\documentclass[review]{elsarticle}

\usepackage{lineno,hyperref}
\modulolinenumbers[5]

\usepackage{amsmath}
\usepackage{amssymb}
\usepackage{amsfonts}
\usepackage{xcolor}
\usepackage{graphicx}
\usepackage{tabularx}
\usepackage{xtab} 		    
\usepackage{multicol}       
\usepackage{booktabs}       

\journal{Chemical Engineering Science}

\begin{document}

\begin{frontmatter}

\title{Semi-supervised Learning for Data-driven Soft-sensing of Biological and Chemical Processes}

\author{Erik Esche}\corref{cor}
\ead{erik.esche@tu-berlin.de}
\author{Torben Talis}
\ead{talis@campus.tu-berlin.de}
\author{Joris Weigert}
\ead{joris.weigert@tu-berlin.de}
\author{Gerardo Brand Rihm}
\ead{g.brandrihm@tu-berlin.de}
\author{Byungjun You}
\ead{byungjun.you@tu-berlin.de}
\author{Christian Hoffmann}
\ead{c.hoffmann@tu-berlin.de}
\author{Jens-Uwe Repke}
\ead{jens-uwe.repke@tu-berlin.de}
\cortext[cor]{Corresponding author}
\address{Technische Universität Berlin, Sekr.~KWT 9, Process Dynamics and Operations Group, Str.~des 17.~Juni 135, D-10623 Berlin, Germany}

\begin{abstract}
Continuously operated (bio-)chemical processes increasingly suffer from external disturbances, such as feed fluctuations or changes in market conditions. 
Product quality often hinges on control of rarely measured concentrations, which are expensive to measure.
Semi-supervised regression is a possible building block and method from machine learning to construct soft-sensors for such infrequently measured states. Using two case studies, i.e., the Williams-Otto process and a bioethanol production process, semi-supervised regression is compared against standard regression to evaluate its merits and its possible scope of application for process control in the (bio-)chemical industry.
\end{abstract}


\end{frontmatter}


\section{Introduction}
Continuously operated chemical and bio-chemical processes are subject to various driving forces, which induce fluctuations \cite{Esche.2020}. Among these are changing compositions of available raw material, demand response mechanisms to balance the power grid, and fluctuations in demand.
Ensuring resource and energy efficiency while at the same time guaranteeing product qualities is a challenging task.
Various methods exist to continuously improve the operation of (bio-)chemical processes by model-predictive control \cite{Rawlings.2009}, non-linear model predictive control with economic objectives \cite{Engell.2007}, or real-time optimization \cite{Camara.2016}.
Especially the latter two require accurate knowledge of the current state of the process to determine suitable modifications to operation trajectories and operation conditions to improve process efficiency.
Despite large advances in recent years, numerous challenges remain \cite{Alexander.2020}. Among these are the availability of suitable models, availability of necessary measurement sensors, the appearance of (structural) plant-model mismatch, and the speed and reliability of the solution of the state estimation problem online.

Regarding (bio-)chemical processes, a recurring issue lies in the availability and frequency of quality measurements. Hence, dedicated research investigated the effect of low sampling rates of important quality measurements and how these measurements can be incorporated into state estimators \cite{Guo.2015, Liang.2009, Hakerl.2018}. By now, this has also been introduced into state estimators applicable to highly nonlinear systems \cite{LopezNegrete.2012} and been amended by filtering of gross errors \cite{Nicholson.2014}. 
However, \cite{Weigert.2018} previously noted that the made assumptions on measurement frequencies in these works are still quite high.
In many (bio-)chemical processes, quality measurements are available every other hour, day, or even week. At the same time, process disturbances affecting product quality may appear in the minute range.

For these and other reasons, fully data-driven softsensing and process monitoring has long been under investigation \cite{Ge.2017} and is gaining traction given the recent advances in machine learning methods. In most cases, these approaches operate without any first principles models and are solely based on measurement data. 
By now, there are abundant examples of process monitoring applications in the chemical industry that range from partial least squares regression \cite{Kresta.1994, Zhang.2009}, over Fisher's discriminant analysis \cite{Zhang.2007, Yu.2011}, and more recently to artificial neural networks \cite{Gonzaga.2009, Iliyas.2013} as well as Gaussian process regression \cite{Grbic.2013, LeZhou.2015}.

While initial applications focused on small sub-processes or single apparatuses, statistical soft-sensors have by now found their way into process control for the steel industry \cite{Kano.2008} or for the quality control of industrial polymerization processes \cite{Gonzaga.2009}. Especially the latter describes frequently arising situations from the perspective of chemical processes: Viscosity of the produced polymers is the essential quality parameter, but measurements can only be obtained \textit{ex situ} and at low frequency. Using the remainder of available measurements to infer the viscosity allows for quality control.
Similarly, \cite{Shen.2020} developed a multi-rate state estimator based on an artificial neural net. 

Construction of such soft-sensors is still a large effort, as available measurement data has to be analyzed manually and its information content regarding the quality parameters to be predicted needs to be evaluated. Usually, only continuously available measurement data is used for such soft-sensors.
Here, we would like to propose a methodology, which allows for a straight forward construction of soft-sensors based on continuous measurement data as well as rarely measured qualities.
Borrowing from recent advances in machine learning, semi-supervised regression (SSR) will be investigated to this end and applied to systems with measurement frequencies and disturbances typical of (bio-)chemical processes in industry.

\subsection{Review of Semi-supervised Learning}

\begin{figure}[th]
	\centering
	\includegraphics[width=0.7\linewidth]{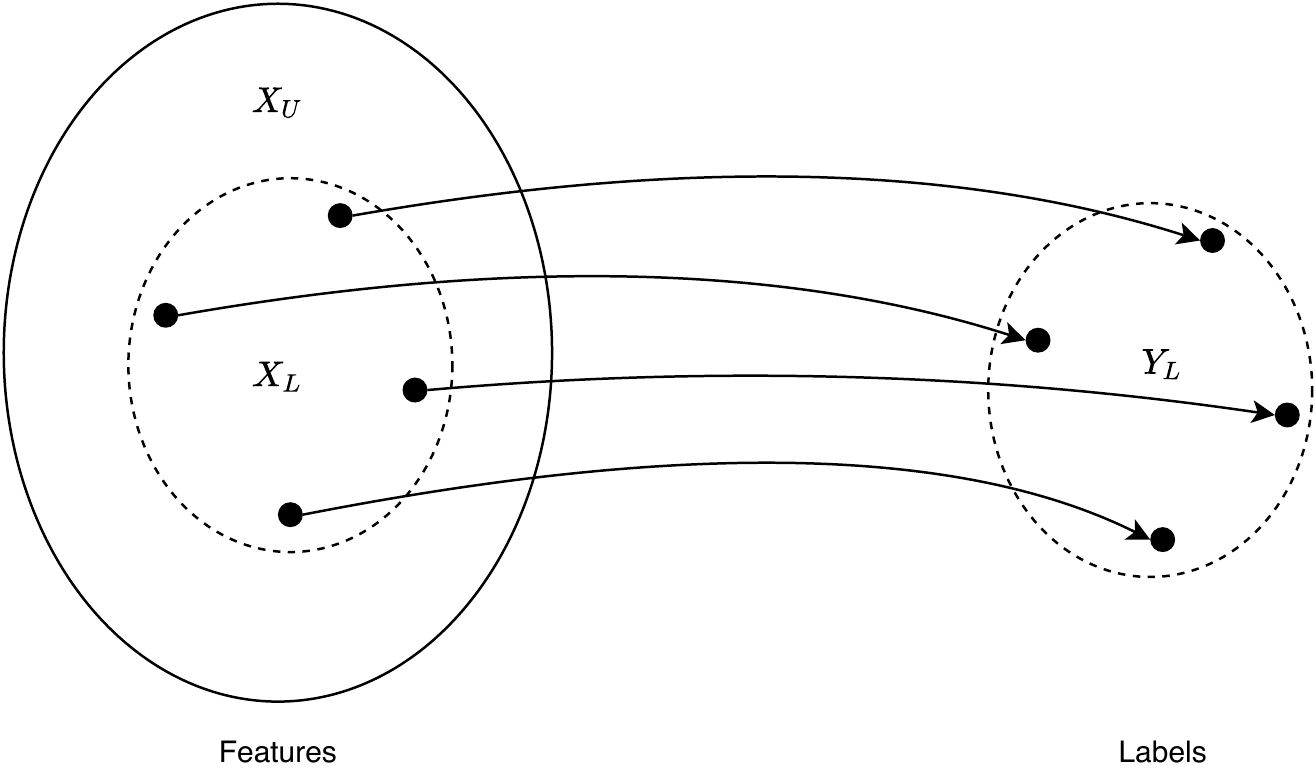}
	\caption[SSR Visualization]{Visualization of the semi-supervised learning problem.}
	\label{fig:01-ssr-problem-visualization}
\end{figure}

In the last 20 years, the area of semi-supervised learning has steadily gained attention \cite{Zhu.2009}. Mostly, this field is concerned with classification, where only few sets of labeled data exist and large amounts of unlabeled data might be used to improve the classification (see Fig.~\ref{fig:01-ssr-problem-visualization}). 

In the following $x$ will be a vector of features and $y$ a label. \textit{From the perspective of chemical processes, $x$ will later be a set of continuous measurements, whereas $y$ will be a rare measurement.}
A set of data points for $x$ combined with respective labels $y$ will be the set of labeled data where $m$ is the number of labeled points: 
\begin{align*}
	\mathcal{L} = \lbrace (x_1, y_1), \ldots, (x_m, y_m) \rbrace
\end{align*}
with the respective set of ``labeled features'' $X_L := \lbrace x_1, \ldots, x_m \rbrace$ and set of available labels $Y_L :=  \lbrace y_1, \ldots, y_m \rbrace$. Further data points without labels will be denoted by $X_U = \lbrace x_{m+1}, \ldots x_{m+n} \rbrace $ with $n$ as the number of unlabeled data. In typical settings, it can be expected that $n \gg m$

A lot of work went into understanding how unlabeled data could be useful for improving learning in cases of few labeled data and how to design algorithms for using unlabeled in addition to labeled data sets \cite{XueyuanZhou.2014}.
A wide variety of models has so far been considered for semi-supervised learning, among these are self-training, mixture models, co-training and multiview learning, graph-based methods, and semi-supervised support vector machines \cite{Zhu.2009}.

All graph-based, semi-supervised classification methods are function estimators and are hence also suitable to perform regression tasks. In most cases so far, Gaussian processes were augmented with specialized kernels constructed on unlabeled data.
In \cite{Lafferty.2000} a first rigorous statistical analysis of semi-supervised regression was carried out to determine how estimators should be constructed to obtain improved predictions for labels in case unlabeled data is also used. They found that a number of initial implementations did not lead to improved convergence rates and sometimes performed worse than regressors using only the available labeled data sets.


Since then, a few first applications with a connection to chemical engineering have been seen: \cite{Zhong.2014} proposed a semi-supervised Fisher discriminant analysis method for monitoring industrial batch processes and identifying faults, which primarily is a classification task.
First attempts at regression are published by \cite{Jin.2012}, who developed a soft-sensor based on principles of semi-supervised regression using irregular and data with missing values.
Analogously, \cite{Ge.2014}  designed a semi-supervised principal component regression model as a soft sensor and \cite{LeZhou.2015b} a semi-supervised probabilistic latent variable regression model for process monitoring.
Similarly, \cite{Yu.2020} recently used semi-supervised hybrid local kernel regression to predict quality parameters of a rubber mixing process for application as a soft sensor using both historical and online data.
However, in the latter four either the regression models were rather simple or the focus was on data series with ``missing values'' -- meaning that $n < m$.

Outside of chemical engineering, numerous recent studies can be found on semi-supervised regression. \cite{Riese.2020} employed self-organizing maps to analyze spectral data from soil samples and to infer moisture. They reported a great improvement compared to existing approaches, e.g., random forest.
\cite{Levatic.2020} employed semi-supervised regression for the chemoinformatics task quantitative structure-activity relationship modeling. Particularly, they highlighted the fact that semi-supervised trees and ensembles thereof show an improved predictive performance compared to their supervised counterparts -- especially with a low number of available labeled data points.

In the meantime, a number of frameworks supporting semi-supervised learning and even semi-supervised regression have become available, e.g., in scikit-learn \cite{FabianPedregosa.2011} with a focus on classification or in TensorFlow \cite{Xie.2019, Jean.2018} specifically for regression tasks.

For the remainder of this contribution, we shall base our work on the implementation described in \cite{Jean.2018}, which was designed for cases with vast amounts of unlabeled data points (3'000 to 2'000'000) and very few available labeled points (50 to 500), which compared to other implementations appears suitable for realistic scenarios for (bio-)chemical processes and will be detailed further below.

By default, the framework implements semi-supervised deep kernel learning, which combines the capabilities of deep neural networks to learn features with the ability of Gaussian processes to capture uncertainty. The neural net serves as an input to the Gaussian process and both are trained simultaneously. For training, a loss function combining both likelihood maximization and variance minimization is employed with weighting factor $\alpha$:
	\begin{align}
		L_{\text{semisup}}(\theta) &= \frac{1}{n} \cdot L_{\text{likelihood}}(\theta) + \frac{\alpha}{m} \cdot L_{\text{variance}}(\theta)\\
		L_{\text{likelihood}}(\theta) &= - \log p(Y_L | X_L, \theta)\\
		L_{\text{variance}}(\theta) &= \sum \limits_{x \in X_U} \textrm{Var}(f(x,\theta))
	\end{align}
Herein, $f$ represents the model (combination of neural net and Gaussian process) and $\theta$ the parameters of both, which need to be trained. Moreover, $n$ and $m$ are the numbers of labeled and unlabeled data points.
$L_{\text{likelihood}}$ serves to minimize to error between $Y_L$ and their predictions by $f$ and hence represents a rather standard regression objective.
On the other hand, $L_{\text{variance}}$ describes the variance of the predictions for the data points $X_U$, for which of course no labels exist. Reducing the variance of these predictions amounts to a posterior regularization.
The hyperparameter $\alpha$ naturally induces a bias-variance trade-off: Large values of $\alpha$ heavily reduce the variance of the predicted $y$ and induces a bias, while small values of $\alpha$ increase the output variance and reduce the bias.
Previous studies have shown a minimum of $L_{\text{semisup}}$ for values of $\alpha$ between 0.1 and 10 balancing variance minimization vs.~bias \cite{Jean.2018}.

\subsection{Objective}
Within the scope of this contribution, we want to go beyond the scope of ``missing data'' and investigate realistic scenarios for (bio-)chemical process applications with infrequent or rare quality measurements, which are essential for quality control.
For chemical processes, quality measurements are often essential for ensuring product quality \cite{Eisen.2020}, but in many cases are not available online and at high measurement frequencies. Advances in process analytics allow for more frequent application of online measurements, e.g., by Raman spectroscopy, but are still limited due to the associated costs.
A large portion of chemical plants still relies on \textit{ex situ} gas chromatography or other analytical methods. These measurements are only available at low frequencies ranging from hourly to weekly measurements depending on the setting. Given the hence large level of uncertainty regarding the current process state, plants are typically operated conservatively to ensure product quality.
Thereof arises the question, how much information can be gathered from these rare measurements and whether a combination with continuous measurement data might allow for online process monitoring and control leading to more efficient process operation.
To the best of our knowledge, this has not been previously investigated and might open up a host of new possibilities for straightforward construction of soft-sensors for process applications.

To this end, this contribution will evaluate the capabilities of semi-supervised regression for the application in (bio-)chemical processes with the long-term goal to easily construct soft-sensors, which may be applied for process control.
Here, this will be attempted on systems with a ``known ground truth'', i.e., detailed process models will be employed to generate artificial measurement data.
The framework by \cite{Xie.2019} will be extended towards time series and the ratio of labeled vs.~unlabeled data points will be evaluated to determine how infrequent measurement data may be to still allow for reasonable inference. The influence of the weighting factor $\alpha$ will similarly be investigated to obtain a guideline for future applications with real-life experimental data.
For comparison of the semi-supervised regression, a basic Gaussian process and a deep kernel learning model will be used.

In the following section, the whole methodology is introduced to evaluate the applicability of semi-supervised regression to process applications with varying degrees of data rarity (i.e., changing frequencies for the quality measurements). Afterwards two case studies are introduced with complex dynamic models serving as ground truth, based on which data is generated, models are trained, and finally evaluated. 

\section{Methodology}
A novel, four step procedure is proposed to evaluate the capabilities of semi-supervised regression for softsensing in (bio-)chemical plants. In the first step, complex dynamic process models are devised, which show a typical stiff and highly nonlinear process behavior and realistic time-delays, points with non-differentiabilities, and include PID control loops.
In the second step, simulations are carried out on these high fidelity models. For each model different operation trajectories are derived as they would appear in real-life processes operated continuously but with external disturbances and load changes:
A long trajectory lasting for more than a month of continuous operation is manually designed, with several set-point changes in the meantime signifying load-changes caused by in- / decreases in demand or raw material supply, overlayed by smaller fluctuations, e.g., in the feed streams. 

For the latter, an amplitude modulated pseudo-random binary signal (APRBS) \cite{Nelles.2001} is employed and added on top of the manually generated trajectories. The size of the step changes and the maximum amplitude of the APRBS signal are chosen to cause a suitably high disturbance of the processes' responses, i.e., the APRBS causes a 3 to 5 \% fluctuation in product qualities and the set-point changes induce dynamic responses of the system, which take up to several hours to recapture steady-state operation. Naturally, this design will induce a bias in the results, but it is nearly impossible to generate fully objective, representative trajectories as these will vary with every single real-life plant. Representative trajectories are given below in Section \ref{s:cases-studies}.\\

In order to be able to handle the generated amount of data on our available computing hardware, the size of the continuously sampled time series is limited to several millions. For training on larger datasets we had insufficient memory (limited to 64 GB).
To investigate a wide range of sample frequencies for the quality measurements, the length of the trajectories is set to 1000 hours (ca.~42 days) with a sampling rate of the ``continuous measurements'' at once per second (1 Hz).
This relates both to realistic measurement frequencies in industrial practice and fits the design of our trajectories and case studies, which show the fastest dynamics in the seconds range. Combined with the length of 1'000 hours, this leads to 3.6 million data points, which fits our memory requirements for continuous datasets.
The described trajectories are fed to the simulation models to obtain data for all state variables (3.6 million data points for each). Based on these data points, quality measurements of interest, e.g., product qualities are down-sampled to different sampling frequencies to represent \textit{ex situ} measurements by gas chromatography. The manual measurements are placed at random along the length of the time window of 1'000 hours to generate different sets of quality measurements: 
\begin{itemize}
	\item Set 1: 10 measurements per 1'000 hours equaling a quality sample once every four days,
	\item Set 2: 100 measurements per 1'000 hours equaling a quality sample once every ten hours,
	\item Set 3: 300 measurements per 1'000 hours equaling a quality sample once every three hours,
	\item Set 4: 1'000 measurements per 1'000 hours equaling a quality sample every hour.
\end{itemize}
The remainder of the measurement data (temperatures, flows, levels, actuators, etc.) are left as is, meaning that for each 3.6 million data points are available. These are combined with each of the four sets of measurement data for the selected quality measurement.

Based on the continuous measurements as inputs (or features $x$) and the quality measurement as output (or label $y$) different models are then trained: a combination of a neural net with two hidden layers and a Gaussian process with the semi-supervised objective function (SSDKL model) as designed in \cite{Jean.2018}, the same with a standard regression objective (deep kernel learning (DKL) model), and a standard Gaussian process (GP) in scikit-learn \cite{FabianPedregosa.2011}. Tab.~\ref{tab:training-scenarios} details all cases, which have been investigated for the training of the different models. 

\begin{table}
	\caption{Overview of options for model training. The variation of $\alpha$ is only applicable to semi-supervised deep kernel learning (SSDKL) and not to deep kernel learning (DKL) with standard objective or Gaussian process regression (GP).} \label{tab:training-scenarios}
	\centering
\begin{tabularx}{\textwidth}{|X|l|p{35mm}|l|X|}
	\hline
	\textbf{Data frequency} & \textbf{APRBS} & \textbf{Structuring of Input} & \textbf{Models} & \textbf{Variation of $\alpha$} \\
	\hline
	Set 1 & yes (Y) & $x_k$ (X) & SSDKL (S)& 0.1 \\
	Set 2 & no (N) & $x_k$ steady-state (XS) & & 1  \\
	Set 3 & & $x_{k, k-1, \ldots, k-4}$ (X5) & & 10 \\
	Set 4 & &  & DKL (D) & \\
	 &  & & GP (G)& \\
	\hline
\end{tabularx}
\end{table}

Each entry of the columns in Tab.~\ref{tab:training-scenarios} is combined with all entries of all other columns -- except for the variation of $\alpha$, which is only of meaning for SSDKL.
The artificial measurement data sets are used with and without the APRBS signal added; the label $y_k$ of a time point $k$ is combined with only the current input $x_k$, or the current input and four values from previous time points, or with steady-state data for $x_k$ only. 
By consequence, 120 different training settings are generated to investigate the information content of the datasets. 
For later analysis a short-hand syntax is used to denote the combinations, e.g., 1-Y-X5-S-10 for Set 1 with APRBS using $x_k$ to $x_{k-4}$ for training of the SSDKL model with $\alpha$ set to 10. The abbreviations are given in brackets in Tab.~\ref{tab:training-scenarios}.
For each model the training is performed ten times, which allows for a limited amount of globalization of the training results, and the best out of the ten in terms of root mean square error (RMSE) is taken for comparison. 

After training, RMSE of all 200 model and training options is computed against the same set of testing data to allow for direct comparison. In addition, the trained models are used to predict the entire trajectory of 1'000 hours and the predicted values of $y$ are compared against the original data for all 3.6 million time points. The latter is performed to obtain a qualitative statement of the usability of the trained models for a possible process control application.

\section{Case Studies} \label{s:cases-studies}
Two (bio-)chemical processes are used to test the capabilities of the semi-supervised regression framework.
The first one is rather small and simple containing only 3 unit operations, but already shows nonlinear dynamic behavior typical of a chemical process. The second consists of a reactor cascade and subsequent product purification and features both biological reactions and thermodynamic separation steps. Hence, it is more representative of (bio-)chemical processes  and also shows a high stiffness ratio.
Both have previously been investigated and detailed models are available. These are further augmented to include dynamic phenomena experienced in real-life plants.
\subsection{Williams-Otto Process}
\begin{figure}
	\centering
	\includegraphics[width=\textwidth]{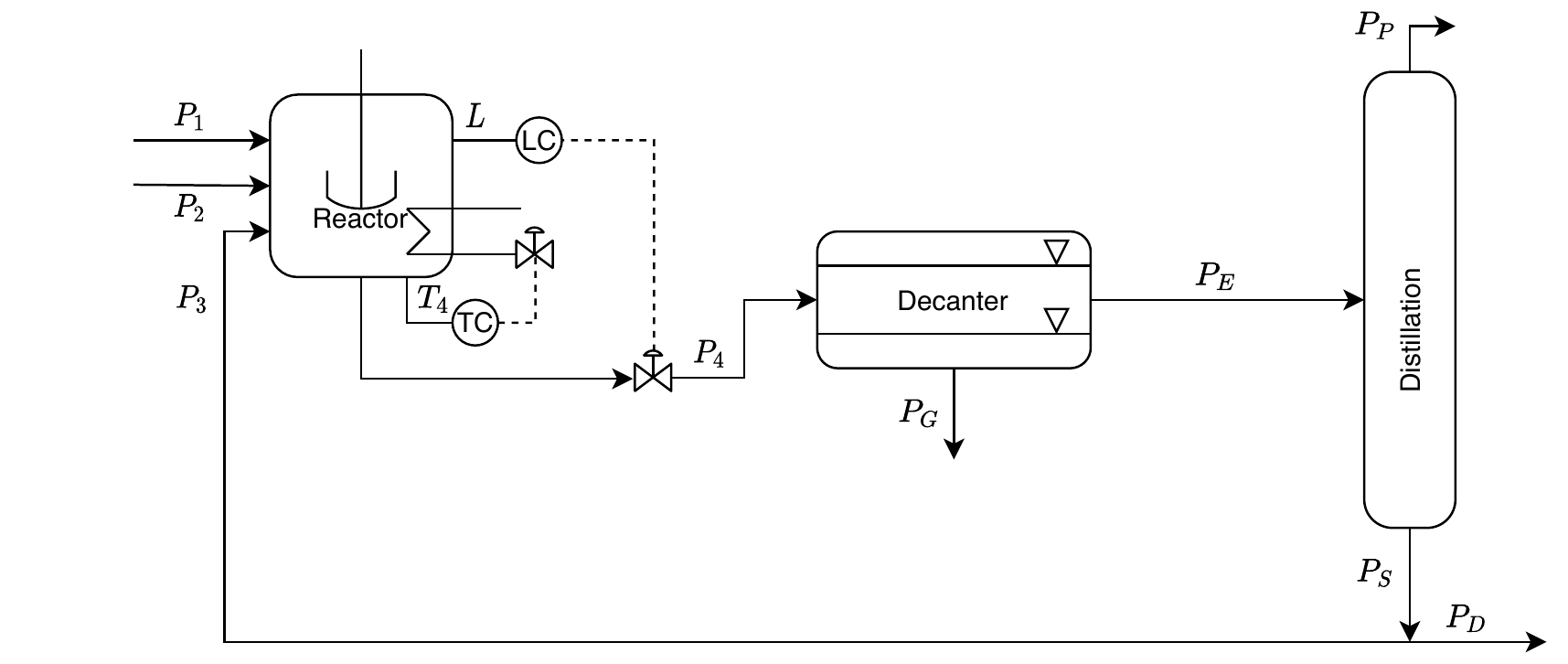}
	\caption{Modified flowsheet of the Williams Otto process.}
	\label{fig:Williams-Otto-Process}
\end{figure}

The well-known Williams-Otto process \cite{Williams.1960} has long been used as a reference process for academic studies and it was originally designed for the ``investigation of computer control''. The process is sketched in Fig.~\ref{fig:Williams-Otto-Process}. In the reactor, component A reacts with B to C and B reacts with C to form product P and byproduct E. In addition, C and P form the unwanted byproduct G, which is removed via phase separation in a decanter. Product P is obtained by the distillate stream of the distillation column while the other components are recycled to the reactor.
A and B are fed to the reactor by streams P1 and P2 separately. The three reactions in the reactor are described using basic Arrhenius type reaction kinetics.
An important constraint for the process operation is the decomposition of the product above 383.15\,K. After cooling in the heat exchanger, some of the byproducts (E and G) are separated in the decanter. The bottom product of the distillation is partially returned to the reactor to further react remaining A and B.\\

In our implementation of the Williams-Otto process, the reactor is assumed to be a continuously stirred tank reactor and described by a set of nonlinear ordinary differential equations describing the mass fractions of the components and the temperature of the reactor. The other unit operations are simplified as ideal separators as in the original implementation by \cite{Williams.1960}.
To bring the dynamic behavior of the model closer to a real-life scenario, time delays accounting for the piping were introduced. With an assumed flow velocity of 2\,m/s and estimated piping length of 10\,m, 8\,m, and 60\,m for the distances from reactor to decanter, decanter to distillation column, and distillation column to the reactor, time delays of 5\,s, 4\,s, and 30\,s were obtained respectively. While the reactor model is implemented in MOSAICmodeling \cite{GregorTolksdorf.2019} and exported to MATLAB, the entire flowsheet including PI controllers and time delays is implemented in MATLAB SIMULINK. The reactor model is available in the Supplementary Material.

With the initial model, the model's responses to set-point changes were examined. Larger steps in feed flows, temperatures, etc., cause a noticeable dynamic response and steady-state operation is regained after 30 to 40 minutes.

\paragraph{Data Selection}
For the further investigations a number of variables are selected, which are typically measured in a chemical plant. Here, those are the flows $F_1$ and $F_2$ relating to the feed streams $P1$ and $P2$ into the reactor, the level $L$ and the temperature $T_4$ of the reactor, flow $F_G$ of the side-stream $PG$, $F_D$ of purge stream $PD$, $F_P$ of product stream $PP$. In addition, the currently implemented heating duty, i.e., actuator value for the heater $Q$, will be assumed available. These variables henceforth constitute the entries of $x$ and will be the starting point for all model training.
For $y$, the fraction of product P in the product stream $PP$ is taken, which is highly relevant for product quality and overall process economics.

\begin{figure}[ht]
	\centering
	\includegraphics[width=\textwidth]{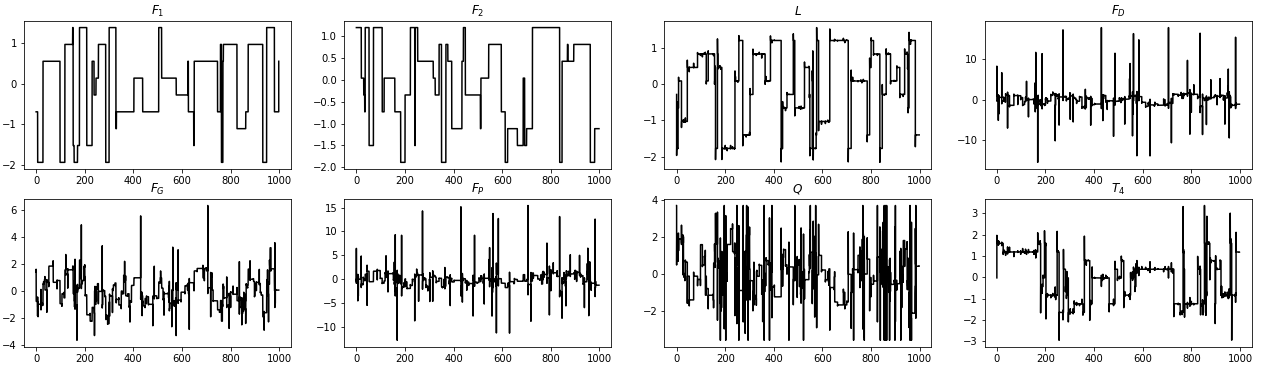}
	\caption{Original data of $x$ for operation trajectory no.~1 for the Williams Otto process (WO-1). All values are normalized. x axis displays operation time in hours.}
	\label{fig:Williams-Otto-Original-Data}
\end{figure}

\begin{figure}[ht]
	\centering
	\includegraphics[width=\textwidth]{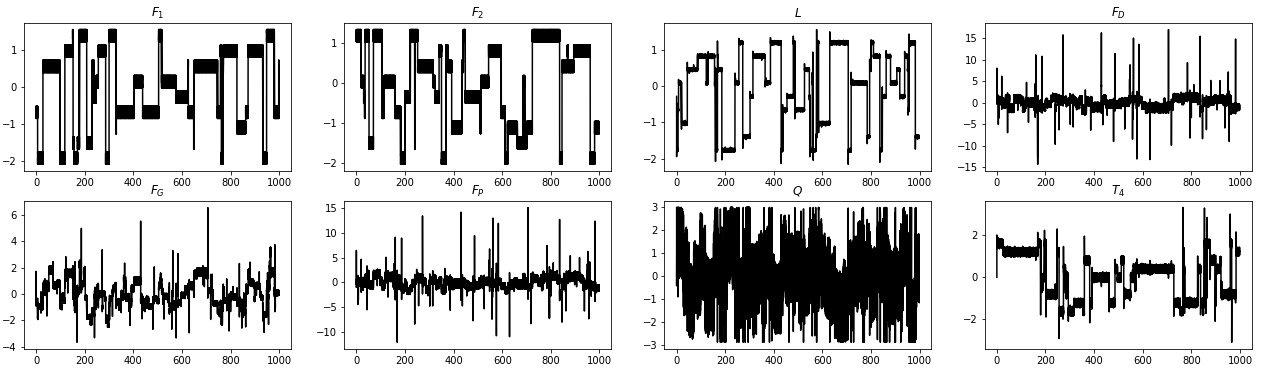}
	\caption{Original data of $x$ for operation trajectory 1 for the Williams Otto process (WO-1) with APRBS signal added to cause a 3.3\% amplitude in the product quality. All values are normalized. x axis displays operation time in hours.}
	\label{fig:Williams-Otto-Original-Data-APRBS}
\end{figure}

\begin{figure}[ht]
	\centering
	\includegraphics[width=\textwidth]{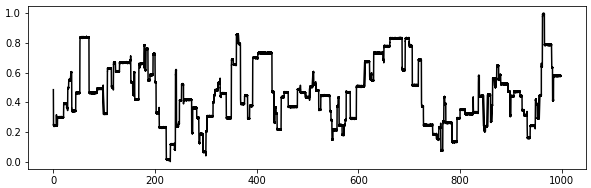}
	\caption{Original data of $y$ for operation trajectory 1 for the Williams Otto process (WO-1) with APRBS signal added to cause a 3.3\% amplitude in the product quality. All values are normalized. x axis displays operation time in hours.}
	\label{fig:Williams-Otto-y-APRBS}
\end{figure}

\begin{figure}[ht]
	\centering
	\includegraphics[width=\textwidth]{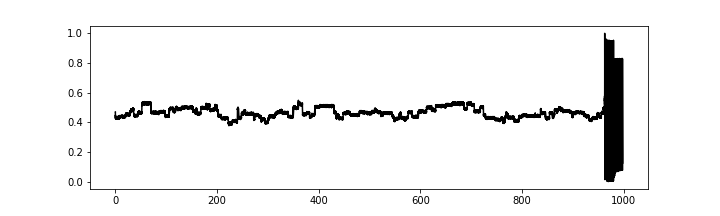}
	\caption{Original data of $y$ for operation trajectory 2 for the Williams Otto process (WO-2) with APRBS signal added to cause a 5\% amplitude in the product quality. All values are normalized. x axis displays operation time in hours.}
	\label{fig:Williams-Otto-2-y-APRBS}
\end{figure}

\paragraph{Operation Trajectories}
Two different operation trajectories, each lasting for 1'000 hours, were generated as described above.
Fig.~\ref{fig:Williams-Otto-Original-Data} displays the original, scaled data for $x$ for the operation trajectory no.~1 of the Williams-Otto process (WO-1) and Fig.~\ref{fig:Williams-Otto-Original-Data-APRBS} the respective trajectories with the added APRBS signal, which is tuned to cause an excitation of $y$ to around 3.3\%. Despite its appearance, the 1'000 hours still contain a number of periods of steady-state.
For the case with APRBS signal added, Fig.~\ref{fig:Williams-Otto-y-APRBS} contains the resulting, normalized profile for $y$. The extent of the induced process dynamics can be seen from the relation between the small 3.3\% excitation and the larger steps, which are one order of magnitude higher than the results of the small excitation.
Additionally, a second operation trajectory has been devised (WO-2), with a larger excitation by the APRBS signal (5\%), which is given in Fig.~\ref{fig:Williams-Otto-2-y-APRBS}. In the final 150 hours of WO-2 the process moves into a highly dynamic operation mode. 

\paragraph{Training and Testing Results}
In the following, representative results of the training and testing systematic will be presented here and discussed in detail. Further results are included in the Supplementary Material. 

Tab.~\ref{tab:results-WO-SSDKLvDKLvGP-lowfreq} compares the results for SSDKL including the variation of $\alpha$ from 0.1 to 10 against the training results for DKL and GP as described above. The respective cell entries are the RMSE regarding the testing data. The percentages relate each SSDKL result to the RMSE of the DKL testing. In Tab.~\ref{tab:results-WO-SSDKLvDKLvGP-lowfreq} we investigate Set 1 with the lowest frequency, i.e., 10 measurements in 1'000 hours. 
The columns belong to the respective two trajectories for the Williams Otto process (WO-1 and WO-2) with different structure configurations for the input data: X, XS, and X5.
A persistent result in all studies is the bad performance of the Gaussian Process (GP) compared to the others whenever the APRBS signal (Y) is added as noise. Of course, we have to remark at this point that no further hyper-parameter tuning was performed up to now and the GP results could possibly be further improved.
Although, it seems improbable that the Gaussian process will see a reduction in the RMSE by orders of magnitued  using this small amount of noisy data available for regression. 
For Set 1 with APRBS, SSDKL consistently outperforms DKL. In Tab.~\ref{tab:results-WO-SSDKLvDKLvGP-lowfreq} the lowest RMSEs per column are set in bold.
Here, the best SSDKL's RMSE is 10 to 50 \% better in terms of RMSE compared to DKL. However, we already see the large influence of $\alpha$ as per column DKL sporadicly outperfoms SSDKL entries.
Regarding the structure of the input data, XS appears to be the best for SSDKL, i.e., $x$ reduced to steady-state data only. The picture is less clear for DKL and GP.
\begin{table}
	\caption{Comparison of SSDKL with variation of $\alpha$, DKL, and GP for Set 1 (lowest frequency) with APRBS of WO-1 and WO-2. Cell entries are RMSE of testing and percentages are relative differences compared to DKL of the respective column. Best results per column are set in bold.} \label{tab:results-WO-SSDKLvDKLvGP-lowfreq}
	\centering
	\begin{tabularx}{\textwidth}{|p{2cm}|X|X|X|X|X|X|}
		\hline
		& \multicolumn{3}{c|}{WO-1} & \multicolumn{3}{c|}{WO-2}\\
		& 1-Y-X & 1-Y-X5 & 1-Y-XS & 1-Y-X & 1-Y-X5 & 1-Y-XS \\
		\hline
		SSDKL & 0.00335 & 0.00372 & 0.00250 & 0.00157  & 0.00197 & \textbf{0.00024} \\
		$\alpha = 0.1$ & +42 \% & +4 \% & -44 \% & -15\% & -1\% & -40\% \\
		\hline SSDKL & \textbf{0.00226}  & \textbf{0.00249}  & \textbf{0.00203}  & \textbf{0.00138}  & \textbf{0.00159} & 0.00029 \\
		$\alpha = 1$ & -13\% & -30\% & -55\% & -25\% & -20\% & -28\% \\
		\hline SSDKL & 0.00520 & 0.00315 & 0.00274 & 0.00162 & 0.00209 & 0.00036 \\
		$\alpha = 10$ & +100 \% & -12 \% & -40 \% & -12 \% & +5 \% & -10 \% \\
		\hline DKL & 0.00260 & 0.00357 & 0.00453 & 0.00185 & 0.00199 & 0.00040 \\
		\hline GP & 0.36534 & 0.49635 & 0.3262 & 0.28866 & 0.45914 & 0.28668 \\
		\hline
	\end{tabularx}
\end{table}

Tab.~\ref{tab:results-WO-SSDKLvDKLvGP-medfreq} shows the same comparison for Set 3, i.e., an intermediate sampling rate with 300 measurements per 1'000 hours.
Similarly, SSDKL shows better results than DKL. Only for a single case (3-Y-X5 for WO-1) DKL shows the best performance. In the rest of the cases, SSDKL is 50 to 80 \% better than DKL regarding the RMSE.
The DKL training has a single goal, which is attaining a good model for the training data. Training of the SSDKL model also emphasizes smoothness through regularization of the predicted $y$ values. Hence, SSDKL will always have a worse fit regarding the training data than DKL. Nevertheless, the regularization ensures that noise from both $x$ and $y$ available for training will be filtered to a certain degree and hence potentially allows for a better match in case of the testing data.

Simlarly to Set 1, this comparison shows the great importance of $\alpha$. Yet again results for different $\alpha$ values can be a lot worse than the DKL results. Interestingly at this higher sampling frequency, the best $\alpha$ value trends to higher values, which is somewhat unexpected.  
It is possible that this is due to the noise available in the training data. With the increased sampling frequency, additional noisy data in $y$ becomes available. With the further increase in $\alpha$ the included noise can be dampened again to ensure a better fit including the testing data also.

Regarding the structuring of the input data, the steady-state only option (XS) still appears to be the best choice here. Showing almost always the lowest RMSE compared to the other structural options for the same training set-up. 

\begin{table}
	\caption{Comparison of SSDKL with variation of $\alpha$, DKL, and GP for Set 3 (intermediate frequency) with APRBS of WO-1 and WO-2. Cell entries are RMSE of testing and percentages are relative differences compared to DKL of the respective column. Best results per column are set in bold.} \label{tab:results-WO-SSDKLvDKLvGP-medfreq}
	\centering
	\begin{tabularx}{\textwidth}{|p{2cm}|X|X|X|X|X|X|}
	\hline
	& \multicolumn{3}{c|}{WO-1} & \multicolumn{3}{c|}{WO-2}\\
	& 3-Y-X & 3-Y-X5 & 3-Y-XS & 3-Y-X & 3-Y-X5 & 3-Y-XS \\
	\hline
	SSDKL & 0.00189 & 0.00230  & 0.00199 & 0.00225 & 0.00426 & 0.00054 \\
	$\alpha = 0.1$ & -66 \% & +390 \% & +60 \% & -38 \% & +31 \% & +145 \% \\
	\hline SSDKL & 0.00094 & 0.00098 & 0.00059 & 0.00105 & 0.00181 & 0.00034 \\
	$\alpha = 1$ & -82 \% & +108 \% & -52 \% & -71 \% & -44 \% & -71 \% \\
	\hline SSDKL & \textbf{0.00075} & 0.00065 & \textbf{0.00058} & \textbf{0.00074} & \textbf{0.00053} & \textbf{0.00022} \\
	$\alpha = 10$ & - 86 \% & +38 \% & -53 \% & -80 \% & -84 \% & -82 \% \\
	\hline DKL & 0.00549  & \textbf{0.00047} & 0.00124 & 0.00361 & 0.00325 & 0.00119 \\
	\hline GP & 0.07394 & 0.42006 & 0.06445 & 0.10051 & 0.36522 & 0.05454 \\
	\hline
\end{tabularx}
\end{table}

Tab.~\ref{tab:results-WO-SSDKL-frequency-alpha} further pursues the effect of $\alpha$ and sampling rate (Set 1 to Set 4) for SSDKL only with structure XS.
As expected, RMSE overall trends to lower values with increasing sampling rates for WO-1, although some exceptions are present.
For WO-2, this is not apparent. Reasons for this can be plentiful. Beyond 850 hours the trajectory moves into an area with intensive dynamics. It is highly possible that this data causes the deterioration of the performance.
Regarding $\alpha$, it is more difficult to draw conclusions based on these initial findings. In general, it appears that RMSE decreases with increasing $\alpha$.

\begin{table}
	\caption{Comparison of the influence of $\alpha$ and sampling frequency (Sets 1 to 4 for WO-1 and WO-2) in terms of the RMSE of the SSDKL model trained using steady-state data only. Entries state the RMSE of each case multiplied by 100.} \label{tab:results-WO-SSDKL-frequency-alpha}
	\centering
	\begin{tabularx}{\textwidth}{|p{1.3cm}|X|X|X|X|X|X|X|X|}
		\hline
		RMSE& \multicolumn{4}{c|}{WO-1} & \multicolumn{4}{c|}{WO-2}\\
		$/ 10^{-2}$ & 1 & 2 & 3 & 4 & 1 & 2 & 3 & 4 \\
		\hline 
		$\alpha=0.1$ & 0.250 & 0.148 & 0.199 & 0.088 & 0.024 & 0.020 & 0.054 & 1.037\\
		$\alpha=1$ & 0.203 & 0.104 & 0.059 & 0.086 & 0.029 & 0.017 & 0.034 & 0.210 \\
		$\alpha=10$ & 0.275 & 0.049 & 0.058 & 0.056 & 0.036 & 0.017 & 0.022 & 0.193 \\
		\hline
	\end{tabularx}
\end{table}
For the remainder of the implemented scenarios with sets 2 and 4 as well as the cases without APRBS (N) the results are highly similar. For the latter, the GP results are better by orders of magnitude compared to the cases with additional noise (Y).

\subsection{Bioethanol Production}
\begin{figure}[ht]
	\centering
	\includegraphics[width=\textwidth]{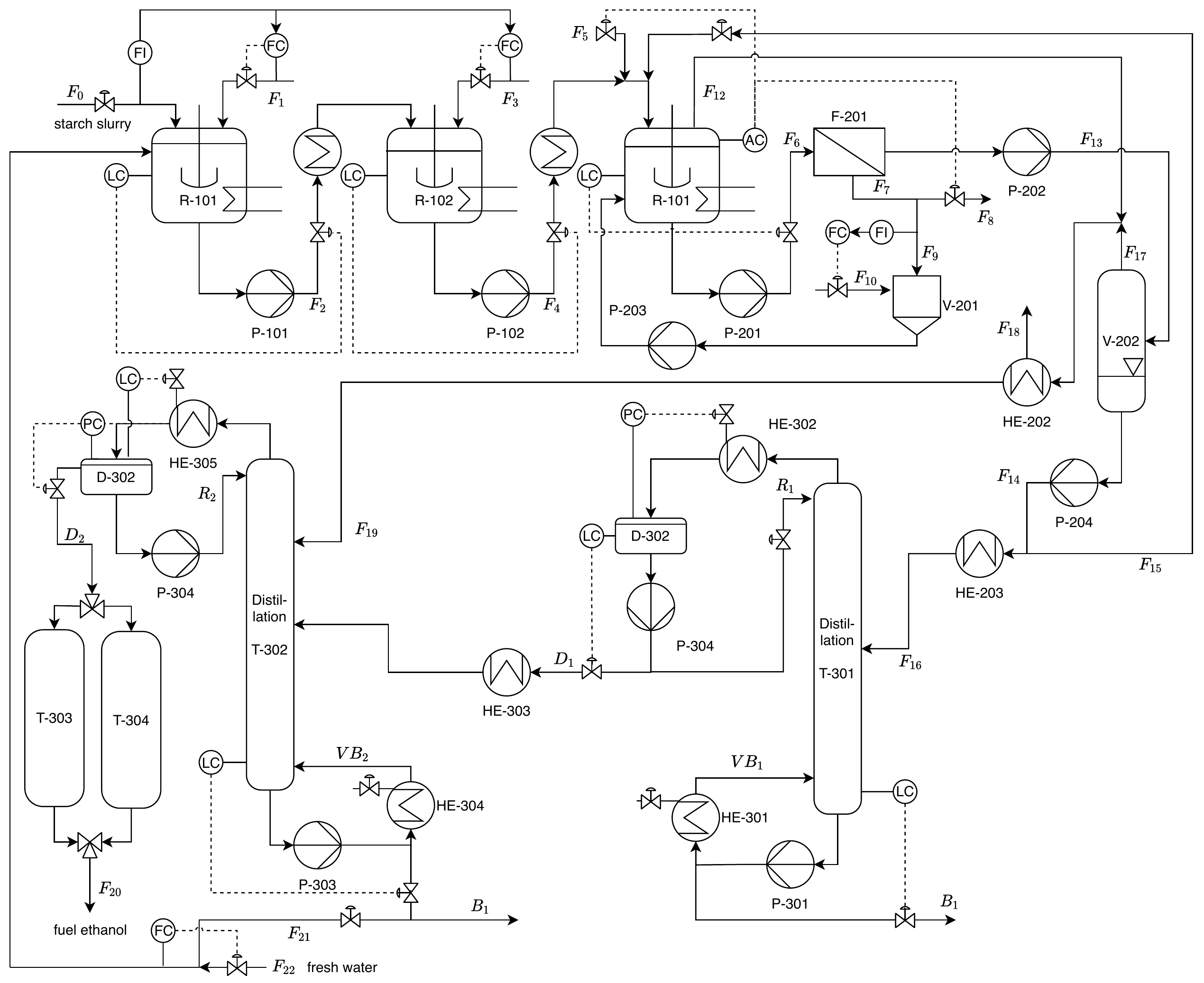}
	\caption{Modified flowsheet of the bioethanol process.}
	\label{fig:Bioethanol-Process}
\end{figure}

The second example has previously been used for case studies on plant-wide optimizing control \cite{SilviaOchoa.2010}. Here, we extend the implementation from \cite{OchoaCaceres.2010}.
The process is displayed in Fig.~\ref{fig:Bioethanol-Process} and describes the production and purification of bioethanol from starch. It consists of three larger sections: an enzymatic starch hydrolysis, a fermentation section, and a purification section dominated by distillation.
The dynamic process model is implemented in MATLAB and MATLAB SIMULINK and described in great detail in \cite{OchoaCaceres.2010}. 
The biological subprocesses are simplified and disregard influence of temperature and pH on the enzymatic activity.

Building on the previously published model a number of modifications and augmentations werde implemented. The manually implemented controllers were replaced by PI controllers of MATLAB SIMULINK and solver ode15s was selected instead of ode14x to speed up computation. As above, time delays were added with an assumed liquid flow velocity of 2\,m/s and 15\,m/s for gas streams. The estimated lengths of the pipes and the resulting time delays are listed in the Supplementary Material.

Compared to the first case study, the model of the bioethanol process is very stiff (stiffness ratio above $10^6$). Depending on which process variable is excited with a step change the overall process regains steady-state after a few hours or up to 300 hours.

\paragraph{Data Selection}
In analogy to the Williams-Otto process, a number of variables are selected here for input data $x$, which are representative of sensors and actuators typically available in a chemical plant. The selection includes the mass flow of the feed stream $F_0$, the reflux flows $R_1$ and $R_2$ to the distillation columns T-301 and T-302, as well as the control valve positions causing the steam flows $VB_1$ and $VB_2$.
For $y$, the economic objective function suggested in \cite{SilviaOchoa.2010} is employed, which is a combination of product quality, energy input, and resource usage. This objective value is directly computed as part of the model.

\begin{figure}[ht]
	\centering
	\includegraphics[width=\textwidth]{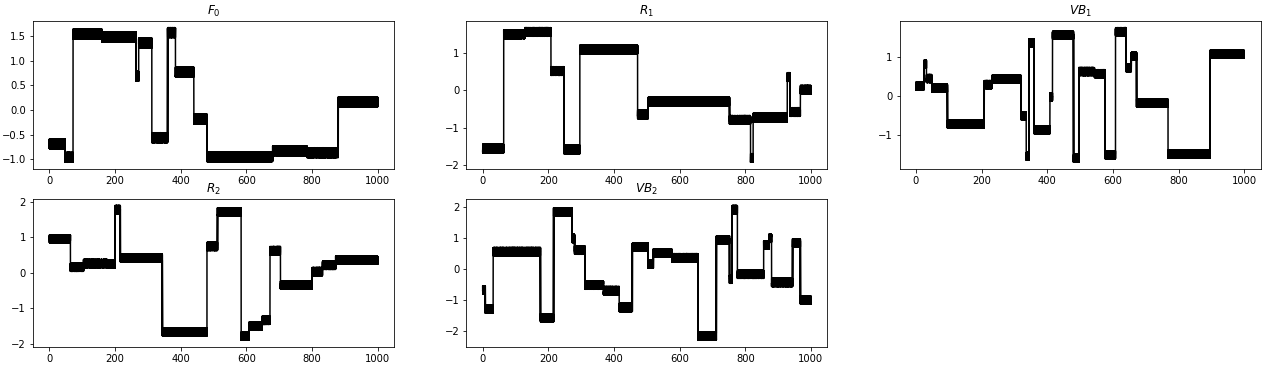}
	\caption{Original data of $x$ for operation trajectory 1 for the Bioethanol process (BE-1). All values are normalized. x axis displays operation time in hours.}
	\label{fig:BioEtOH-Original-Data}
\end{figure}

\begin{figure}[ht]
	\centering
	\includegraphics[width=\textwidth]{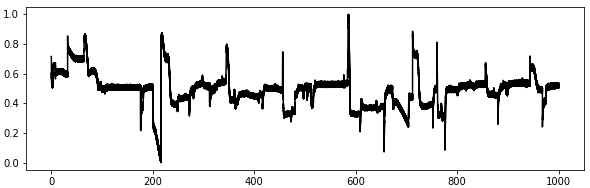}
	\caption{Original data of $y$ for operation trajectory 1 for the Bioethanol process (BE-1). All values are normalized. x axis displays operation time in hours.}
	\label{fig:BioEthOH-y-APRBS}
\end{figure}

\paragraph{Operation Trajectories}
Fig.~\ref{fig:BioEtOH-Original-Data} displays the original data for $x$ for the operation trajectory no.~1 of the Bioethanol process (BE-1). This already includes the added APRBS signal, which is tuned to cause an excitation of $y$ to around 5\%.
Fig.~\ref{fig:BioEthOH-y-APRBS} contains the resulting, normalized profile for $y$.
Additionally, a second operation trajectory (BE-2) has been devised with similar properties but different step changes.

\paragraph{Training and Testing Results}
The presentation of the results is handled as before. Representative results are discussed and further data is included in the Supplementary Material.
In Tab.~\ref{tab:results-BE-SSDKLvDKLvGP-lowfreq} the low frequency results are displayed for all model types. 
Yet again, all structural options for both datasets BE-1 and BE-2 are included. For SSDKL, $\alpha$ is varied from 0.1 to 10 once more.
Similar to the Williams-Otto process example, DKL is in general still outperformed by SSDKL with a sole exception for the steady-state data of BE-1. However, for these, RMSE is still in the same order of magnitude and overall small.
Nevertheless, the difference between SSDKL and DKL is less pronounced here. In most cases the best RMSE of SSDKL is only 5 to 25 \% below the results for DKL.
Structurewise, the steady-state option for the input data still appears preferrable showing consistently lower RMSE values compared to the others.
The disadvantage of the Gaussian process regression also persists. 
\begin{table}
	\caption{Comparison of SSDKL with variation of $\alpha$, DKL, and GP for Set 1 (lowest frequency) with APRBS of BE-1 and BE-2. Cell entries are RMSE of testing and percentages are relative differences compared to DKL of the respective column. Best results per column are set in bold.} \label{tab:results-BE-SSDKLvDKLvGP-lowfreq}
	\centering
	\begin{tabularx}{\textwidth}{|p{2cm}|X|X|X|X|X|X|}
		\hline
		& \multicolumn{3}{c|}{BE-1} & \multicolumn{3}{c|}{BE-2}\\
		& 1-Y-X & 1-Y-X5 & 1-Y-XS & 1-Y-X & 1-Y-X5 & 1-Y-XS \\
		\hline
		SSDKL (0.1) & 0.01114 & 0.0122 & 0.00828 & 0.01128 & \textbf{0.0123} & \textbf{0.00938} \\
		& -22 \% & -5  \% & +35 \% & -10 \% & -10 \% & -13 \% \\
		\hline SSDKL (1) & 0.01172 & \textbf{0.01189} & 0.00758 & \textbf{0.01089} & 0.01301 & 0.01001 \\
		& -18 \% & -7 \% & +23 \% & -13 \% & -6 \% & -7 \% \\
		\hline SSDKL (10) & \textbf{0.01103} & 0.01273 & 0.00722 & 0.01181 & 0.01243 & 0.01108 \\
		& -23 \% & -1 \% & +17 \% & -6 \% & -10 \% & +2 \% \\
		\hline DKL & 0.01424 & 0.01281 & \textbf{0.00615} & 0.01258 & 0.01378 & 0.01082 \\
		\hline GP & 0.26433 & 0.38818 & 0.28064 & 0.24022 & 0.41288 & 0.24696 \\
		\hline
	\end{tabularx}
\end{table}

Moving to the higher frequency, i.e., Set 3 with 300 measurements per 1'000 hours, see Tab.~\ref{tab:results-BE-SSDKLvDKLvGP-medfreq}, the situation does not change a lot. Here also, SSDKL shows better results than DKL. Identically to the Williams-Otto results, the best results are found with increasing $\alpha$ values at this higher frequency (moving from Set 1 to Set 3).
Yet again, we assume that this behavior may be explained as follows: The increase in available noisy data for $y$ with matching noisy data in $x$ requires a stronger regularization to ensure an overall better fit regarding the overall testing data. 

The difference between SSDKL and DKL is still pronounced with RMSE values for SSDKL lower by 8 to 37 \%.
Strikingly, the advantage of the steady-state option increases heavily. The RMSE of XS compared to X and X5 are smaller by up to 50 \%.

\begin{table}
	\caption{Comparison of SSDKL with variation of $\alpha$, DKL, and GP for Set 3 (intermediate frequency) with APRBS of BE-1 and BE-2. Cell entries are RMSE of testing and percentages are relative differences compared to DKL of the respective column. Best results per column are set in bold.} \label{tab:results-BE-SSDKLvDKLvGP-medfreq}
	\centering
	\begin{tabularx}{\textwidth}{|p{2cm}|X|X|X|X|X|X|}
		\hline
		& \multicolumn{3}{c|}{BE-1} & \multicolumn{3}{c|}{BE-2}\\
		& 3-Y-X & 3-Y-X5 & 3-Y-XS & 3-Y-X & 3-Y-X5 & 3-Y-XS \\
		\hline
		SSDKL (0.1) & 0.00464 & 0.00609 & 0.00253 & 0.00513 & 0.00618 & 0.00326 \\
		& -13 \% & -11 \% & +4 \% & -27 \% & -16 \% & -11 \% \\
		\hline SSDKL (1) & \textbf{0.00461} & 0.00497 & 0.00248 & 0.00519 & 0.00571 & 0.00313 \\
		& -14 \% & -28 \% & +2 \% & -26 \% & -22 \% & -15 \% \\
		\hline SSDKL (10) & 0.00477 & \textbf{0.00462} & \textbf{0.00225} & \textbf{0.00477} & \textbf{0.00464} & \textbf{0.00266} \\
		& -11 \% & -33 \% & -8 \% & -32 \% & -37 \% & -28 \% \\
		\hline DKL & 0.00536 & 0.00686 & 0.00243 & 0.00705 & 0.00733 & 0.00368 \\
		\hline GP & 0.18374 & 0.10476 & 0.22637 & 0.18500 & 0.09945 & 0.16935 \\
		\hline
	\end{tabularx}
\end{table}

Tab.~\ref{tab:results-BE-SSDKL-frequency-alpha} repeats the investigation of the influence of $\alpha$ and of the measurement frequency on the RMSE of the SSDKL results. Here also, we limit ourselves to the results for the XS structure.
The situation regarding the influence of $\alpha$ is still not perfectly clear, but the overall trend persists: A reduction of the RMSE can be achieved by increasing $\alpha$ to 10. Certainly, this value needs to be tuned on a case-by-case basis. 
In opposition to the Williams-Otto examples, the influence of an increase in the measurement frequency is clearer here. Going from Set 1 to Set 4, the RMSE steadily decreases as expected. 
	
\begin{table}
	\caption{Comparison of the influence of $\alpha$ and sampling frequency (Sets 1 to 4 for BE-1 and BE-2) in terms of the RMSE of the SSDKL model trained using steady-state data only. Entries state the RMSE of each case multiplied by 100.} \label{tab:results-BE-SSDKL-frequency-alpha}
	\centering
	\begin{tabularx}{\textwidth}{|p{1.3cm}|X|X|X|X|X|X|X|X|}
		\hline
		RMSE & \multicolumn{4}{c|}{BE-1} & \multicolumn{4}{c|}{BE-1}\\
		$/ 10^{-2}$ & 1 & 2 & 3 & 4 & 1 & 2 & 3 & 4 \\
		\hline 
		$\alpha=0.1$ & 0.828 & 0.354 & 0.253 & 0.180 & 0.938 & 0.492 & 0.326 & 0.276\\
		$\alpha=1$ & 0.758 & 0.329 & 0.248 & 0.216 & 1.001 & 0.593 & 0.313 & 0.233\\
		$\alpha=10$ & 0.722 & 0.365 & 0.225 & 0.165 & 1.108 & 0.425 & 0.266 & 0.202 \\
		\hline
	\end{tabularx}
\end{table}

\subsection{Evaluation and Analysis}
To employ soft-sensors for process control, their predictions need to match steady-state data as well as mimic dynamic behavior, which need to be countered by control actions.
The discussion of the results so far has solely looked at the RMSE regarding the testing data, which of course does not mean that all dynamics are caught although they might be significant for control.
Considering the information content included in $y$ in sets 1 to 4, correctly predicting dynamics is a tall order.
Based on the results, choice XS was always the best for training. This signals  trouble regarding the desired prediction of dynamics:
The steady-state data and especially the little data on $y$ does not contain information about dynamics and how the dynamics in $x$ should be propagated to $y$ without further information is not obvious.

Regarding the RMSE, the results of the two case studies appear as expected. The posterior regularization controlled by $\alpha$ shows a positive effect regarding the testing data and the set-up is surprisingly well able to handle noise in spite of the results for GP. 

\begin{figure}[ht]
	\centering
	\includegraphics[width=\textwidth]{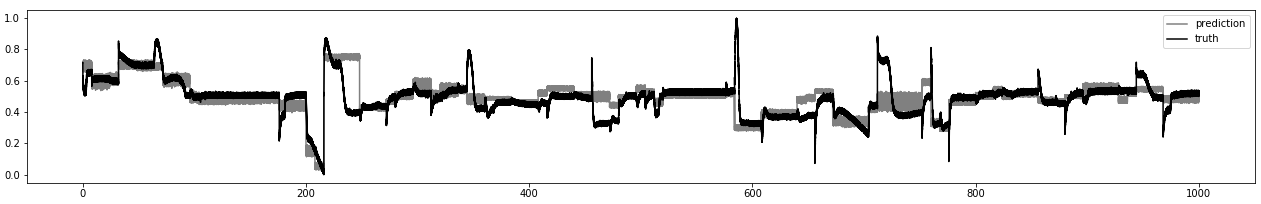}
	\caption{Comparison of prediction and truth of $y$ for BE-1 over the entire time horizon of 1'000 hours. The prediction is based on the SSDKL model obtained from Set 2, steady-state data only and $\alpha$ = 10.}
	\label{fig:BioEtOH-Dynamic Profile-1000}
\end{figure}

To further investigate the prediction of process dynamcis, Fig.~\ref{fig:BioEtOH-Dynamic Profile-1000} presents a comparison of the original operation trajectory BE-1 for $y$ (``truth'') and its predicted counterpart (``prediction''). The prediction is built using the SSDKL model obtained with $\alpha=10$, steady-state data only and then applied on the entire input trajectory in $x$.

\begin{figure}[ht]
	\centering
	\includegraphics[width=\textwidth]{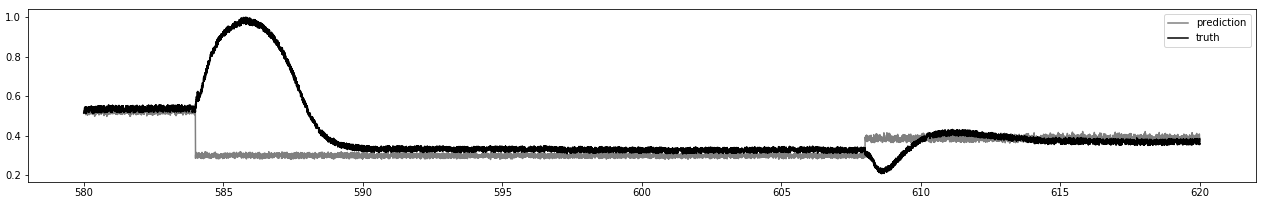}
	\caption{Comparison of prediction and truth of $y$ for BE-1 for a reduced interval from operation time 580 to 620 hours. It is obvious that the prediction can not match the dynamics shown in the actual data (``truth''). To also include dynamics in the SSDKL prediction further data would be necessary or a combination with another method, which is able to mimic the dynamic trends of the underlying process.}
	\label{fig:BioEtOH-Dynamic Profile-580-620}
\end{figure}

As can be seen, \textit{prediction} and \textit{truth} show a larger match for ranges with steady-state behavior. Whenever there are stronger dynamics in the original profile, the two deviate strongly.
This is even more apparent in Fig.~\ref{fig:BioEtOH-Dynamic Profile-580-620}, which shows an excerpt from operation hours 580 to 620. Here, the steady-state match is almost perfect, while the dynamics cannot be caught by the prediction for $y$.
With the available model structures and training scheme, it appears that only two possibilities remain to close these gaps and match the underlying process dynamics: (1) to ensure a higher data density in $y$ or (2) to combined the SSDKL model with a different technique, which correctly predicts dynamics, but is unable to correctly match the steady-state values.
The former could of course only be achieved by online process analytics and imply a costly solution.
For the latter, possible combinations with dynamic surrogate models should be further investigated, but are beyond the scope of this contribution.

The results so far signal that while the steady-state prediction can probably be achieved and that SSDKL appears to be a helpful building block for constructing soft-sensors for operation of (bio-)chemical processes, an accurate prediction of process dynamics is not realistic.
Regarding the steady-state, there are of course also some deficiencies, i.e., smaller offsets whenever the process moves into areas where no measurement data for $y$ has yet been provided. 
As can be seen from Tab.~\ref{tab:results-BE-SSDKL-frequency-alpha} the measurement frequency does play a meaningful role regarding the accuracy of the predictions. For the scenarios WO-1 / -2 and BE-1 / -2 devised here, measurements every 10 hours (Set 2) and more appear reasonable, while the prediction error for Set 1 grows to more than 10 \% at times.
Of course, from a real-life perspective this question has to be answered with a slightly different perspective also: How long is the time horizon of the plant, in which no fundamental changes were implemented, so how much data is available for training of the SSDKL model. Secondly, are there time-dependent deteriorations, such as catalyst deactivation or fouling, which need to be considered also. Finally, how large is the prediction error of $y$ allowed to be with respect to the economics of the respective plant.

\section{Conclusions and Outlook}
Constructing soft-sensors for (bio-)chemical processes still is an arduous and time-consuming task. The usage of expensive and infrequent measurements is difficult in practice. In the present contribution we have investigated whether semi-supervised learning can serve as a facilitator for improved prediction of process states, which are seldom measured. With the two case studies of the Williams-Otto and a bioethanol production process, we have shown that even in settings with very few measurements, i.e., every 100 hours or every 4 days, predictions of intermediate states is possible with reasonable error.

Compared to standard, supervised regression (here DKL and GP), the adjusted training scheme shows a reduction in terms of the RMSE with respect to testing data. There is, however, a high sensitivity regarding the hyper-parameter $\alpha$. It appears based on the present data that for further applications $\alpha$ has to be tuned by validation experiments.

Naturally, our evaluation here was solely based on artificially generated experimental data. As a next step, the whole scheme needs to be translated to real-life process applications. Furthermore, to reach the long-term goal of self-optimizing process control, a further combination of an SSDKL-based predictor with a dynamic process model should be investigated. 

\section*{Acknowledgements}
The authors would like to thank the Max Buchner Foundation for their financial support (MBFSt-Kennziffer: 3733).

\section*{Nomenclature}
\begin{multicols}{2}
	\begin{xtabular}{c>{\raggedright\arraybackslash}p{0.35\textwidth}}
		\textbf{Abbreviations} & \\
		APRBS & Amplitude modulated pseudo-random binary signal\\
		BE-1 & Trajectory no.~1 for the bioethanol process\\
		BE-2 & Trajectory no.~2 for the bioethanol process\\
		DKL & Deep-kernel learning\\
		GP & Gaussian process\\
		RMSE & Root mean square error\\
		SSDKL & Semi-supervised deep-kernel learning\\
		SSR & Semi-supervised regression\\
		WO-1 & Trajectory no.~1 for the Williams-Otto process\\
		WO-2 & Trajectory no.~2 for the Williams-Otto process\\
		\textbf{Greek Symbols} & \\
		$\alpha$ & weighting factor\\
		$\theta$ & model parameters\\
		\textbf{Latin Symbols} & \\
		D & short for DKL\\
		G & short for GP\\
		$L$ & objective function for training\\
		$\mathbb{L}$ & set of labeled data\\
		$m$ & number of labeled points\\
		$n$ & number of unlabeled points\\
		S & short for SSDKL\\
		$x$ & input data / features \\
		$X$ & set of $x$\\
		X & input structure using only one time point\\
		X5 & input structure using 5 time points\\
		$y$ & output data / labels \\
		$Y$ & set of $y$\\
		Y & including APRBS\\
		\textbf{Indices} & \\
		$k$ & discrete time points\\ 
		\textbf{Subscripts} & \\
		$L$ & labeled\\
		N & without APRBS\\
		semisup & semi-supervised\\
		$U$ & unlabeled\\
	\end{xtabular}
\end{multicols}

\bibliography{mybibfile}

\end{document}